          [1999/12/02 1.01 (PWD)]
\documentclass[a4paper,twocolumn]{esapub} 

\def\ltsima{$\; \buildrel < \over \sim \;$}
\def\lsim{\lower.5ex\hbox{\ltsima}}
\def\gtsima{$\; \buildrel > \over \sim \;$}
\def\gsim{\lower.5ex\hbox{\gtsima}}

\usepackage{epsfig}

\title{Derivation of preliminary IBIS response matrices with the INTEGRAL Simulator}
\author[1,2]{D. G\"{o}tz}
\author[1]{D.I. Cremonesi}
\author[1]{S. Mereghetti}

\affil[1]{Istituto di Fisica Cosmica ``G.Occhialini'' -- CNR,
Milano, Italy}
\affil[2]{Dipartimento di Fisica -- Universit\`a degli Studi di Milano, Italy}

\begin{document}


\maketitle

\begin{abstract}

We have used the IBIS Simulator to produce preliminary response
matrices for the ISGRI and PICsIT detectors
in order to  help understanding their scientific performances
before the calibration results are available.
The derived matrices, in a format compatible with the  XSPEC spectral analysis package,
have been tested by fitting simple models
and then used to analyze simulations of
astrophysical sources with more complex spectra.

\end{abstract}

\section{The Simulator}

The IBIS Simulator (developed in collaboration by
IFC/CNR Milan, Southampton University and ISDC) can perform
the following steps.
It first generates a model of the gamma-ray sky with the possibility
of defining position, intensity and spectral properties
of the celestial sources. Then  different
observation strategies (pointing directions, observation length, dithering
patterns, etc.) can be defined.
After these preliminary steps the interaction of all the source
photons with the
active and passive materials of the instruments are simulated.
An estimate of the  background is also added, based on the   results
of the INTEGRAL
Mass Model (Lei et al. 1999), which
is a detailed Monte Carlo simulation that takes into account
the interaction of the particle and photon  flux with the whole INTEGRAL spacecraft.
The Simulator finally produces a data set in a format fully compatible
with the ISDC analysis system.
For the analysis
of the output data we have developed some prototype programs,
which perform image deconvolution and spectral extraction.

\section{Building the response matrices}

\begin{figure}[ht!] 
\centerline{\epsfig{file=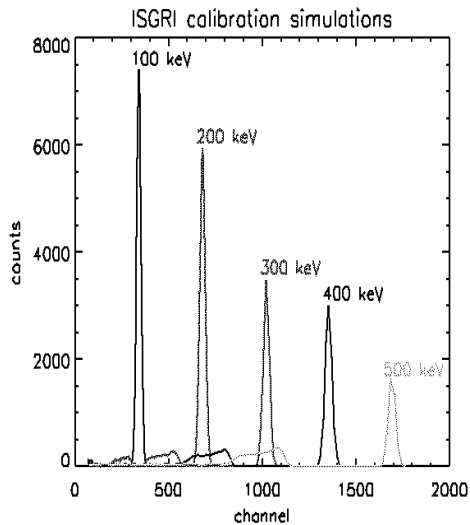,height=3.4in,width=3.5in,angle=-90}}
\vspace{10pt}
\caption{Simulated ISGRI response to on-axis mono-energetic sources of
different energies. The spectra are binned with different factors.}
\label{ISGRI}
\end{figure}

\begin{figure}[ht!] 
\centerline{\epsfig{file=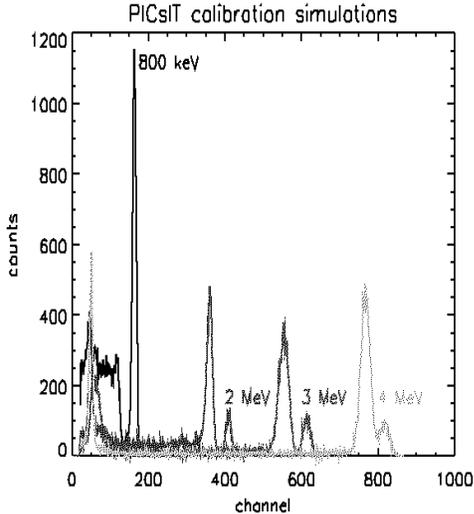,height=3.4in,width=3.5in,angle=-90}}
\vspace{10pt}
\caption{Simulated PICsIT response to on-axis mono-energetic sources of
different energies.}
\label{PICsIT}
\end{figure}

To build the response matrices of ISGRI and PICsIT (Ubertini et al. 1999),
two sets of simulations of on-axis monochromatic sources have
been performed at different energies (20 values for ISGRI between
15 keV and 600 keV and 18 for PICsIT between 100 keV and 5 MeV).
An example of the resulting data for some sample energies can be seen in
Figures \ref{ISGRI} and \ref{PICsIT}.

\begin{figure}[ht!] 
\centerline{\epsfig{file=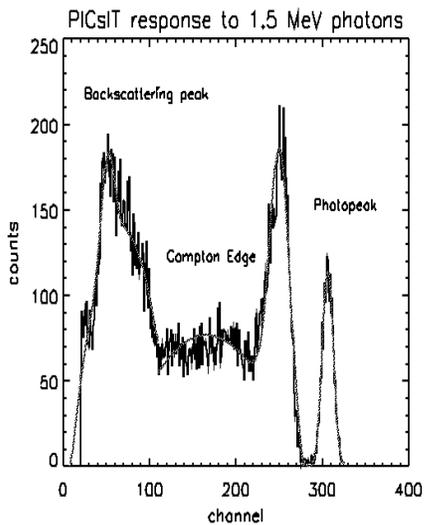,height=3.4in,width=3.5in,angle=-90}}
\vspace{10pt}
\caption{Simulated PICsIT response to a mono-chromatic source of 1.5 MeV photons.}
\label{PICsIT2}
\end{figure}

The simulations clearly show
the photopeak, the Compton edge and, in the case of PICsIT the backscattering
peak (see Figure \ref{PICsIT2}). These three features have been
fitted separately with analytical functions. The photopeak required a Gaussian
function, while the other two required a Gaussian plus a
quadratic function.
%

\begin{figure}[ht!] 
\centerline{\epsfig{file=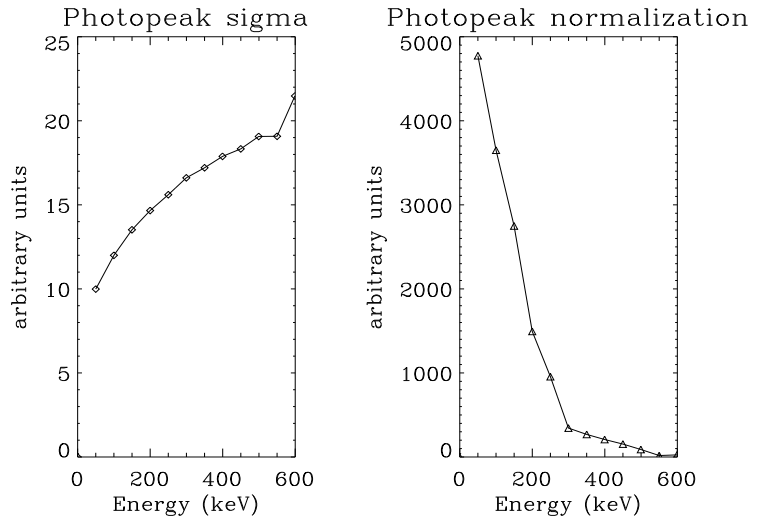,height=4in,width=2.8in,angle=90}}
\vspace{10pt}
\caption{Interpolation of the photopeak parameters of the ISGRI response.}
\label{par}
\end{figure}

The resulting set of parameters as a function of the energy
have been then interpolated over the
entire range of energies of the two instruments (see Figure \ref{par}).

Finally these results have been used to create two FITS files (for
each detector): an RMF file containing the normalized matrix and
an ARF file which contains the information about the efficiency of
the instruments. These two files have been created in such a way
that they are compatible with the XSPEC data analysis package
(OGIP standard ver. 1992a).\\



\begin{figure}[ht!] 
\centerline{\epsfig{file=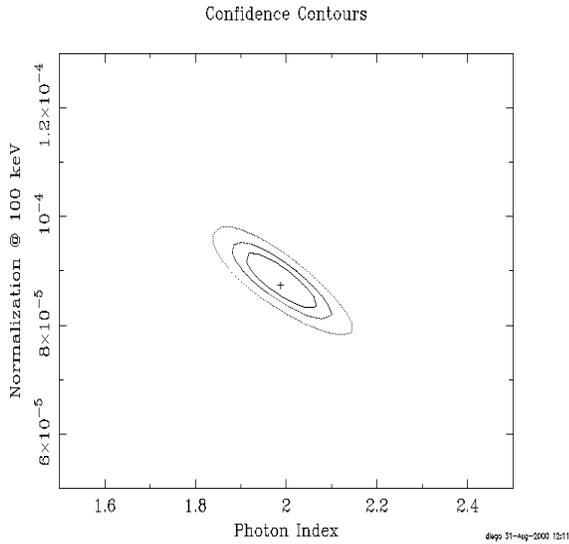,height=3.1in,width=3.1in,angle=-90}}
\vspace{10pt} \caption{Confidence contours of the power law parameters for a $\sim$1000 s
observation of a $\sim$100mCrab source.} \label{cont}
\end{figure}

The matrices have been tested with different input spectra. An
example is the following: a $\sim$100mCrab, on-axis
source with a power law spectrum observed for 1050 s
(input parameters: photon index $\Gamma$=2,
A$_{100 keV}$ = 9$\times$$10^{-5}$ ph cm$^{-2}$ s$^{-1}$  keV$^{-1}$).
The best fit with XSPEC yielded
$\Gamma$ = 1.987 $\pm$ 0.0483 ,
A$_{100 keV}$ = (8.744 $\pm$ 0.6357)$\times$$10^{-5}$ ph cm$^{-2}$ s$^{-1}$  keV$^{-1}$,
Reduced $\chi^{2}$ = 0.942394.
The confidence contours of the fit are presented in Figure \ref{cont}.

\section{Scientific applications}

\subsection{Nova Muscae (GRS 1124-684)}

\begin{figure}[ht!] 
\centerline{
\epsfig{file=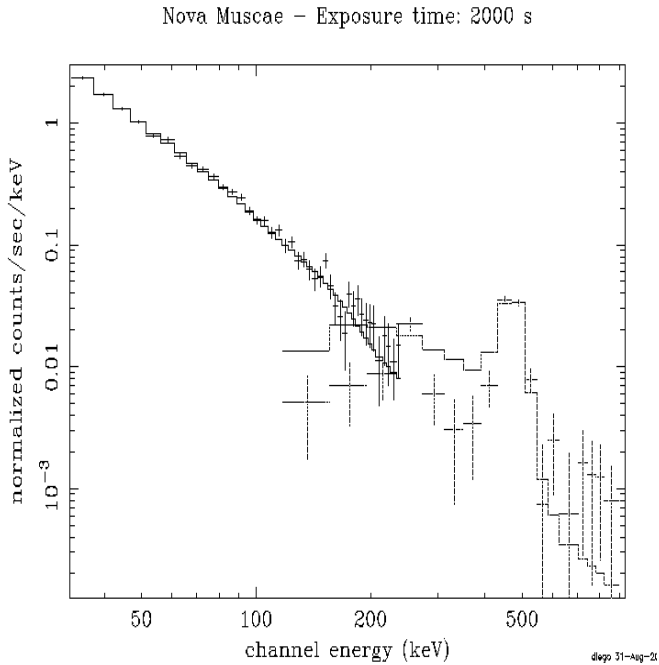,height=3.8in,width=3.8in,angle=-90}}
\vspace{10pt} \caption{2000 s IBIS simulation of Nova Muscae.} \label{nm}
\end{figure}

The X-ray transient GRS 1124-684 (Nova Muscae) was discovered in 1991
with the  GRANAT  and GINGA satellites. Measurement of its
mass function during quiescence established the presence of
a black hole (Remillard et al. 1992).
GRS 1124-684  was observed several times by the SIGMA telescope
during its outburst (Goldwurm et al. 1992). An emission feature
at $\sim$500 keV was discovered during the last 13 hours of the
January 20th observation.

We have simulated IBIS observations of this spectral feature,
with exposure times ranging from 125 s to 15,000 s, using as input
parameters the values measured by SIGMA.
An example of a 2000 s
simulation (see Table \ref{table1}) is shown in Figure \ref{nm}. The simulations show
that the $\sim$500 keV feature can be
appreciated ($\sim$3 $\sigma$) for observations as short as 500 s. 

\begin{table*}[ht!]
\caption{IBIS simulation of Nova Muscae - Exposure time: 2000 s.}
\label{table1}
\begin{center}
\begin{tabular}{|c|c|c|}
\hline \hline
\ & Input parameter & Best fit result \\
\hline
&&\\
Photon Index & 2.42  & 2.423 $\pm$ 0.0184 \\
&&\\
Norm. at 1 keV & 11.27 ph cm$^{-2}$ s$^{-1}$ keV$^{-1}$ & 10.68 $\pm$ 0.741 ph cm$^{-2}$ s$^{-1}$ keV$^{-1}$ \\
&&\\
Line energy & 481 keV & 483.3 $\pm$ 2 keV \\
&&\\
Line width ($\sigma$) & 23 keV & 20.66 $\pm$ 0.1199 keV \\
&&\\
Line flux & 6$\times$$10^{-3}$ ph cm$^{-2}$ s$^{-1}$ & (3.017 $\pm$
0.242)$\times$$10^{-3}$ ph cm$^{-2}$ s$^{-1}$ \\
\hline
\hline
\end{tabular}
\end{center}
\end{table*}

\subsection{IBIS line sensitivity}
\label{IBIS line sensitivity}

\begin{figure}[ht!] 
\centerline{\epsfig{file=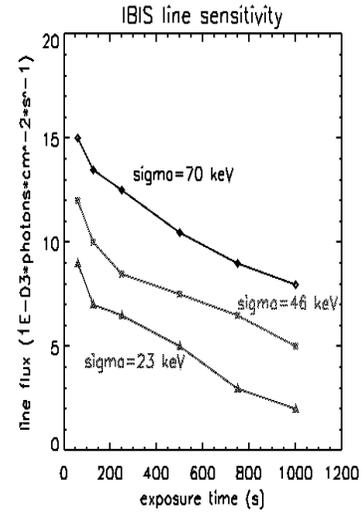,height=3.2in,width=3.4in,angle=-90}}
\vspace{10pt}
\caption{IBIS line sensitivity obtained from the simulations described in
Section \ref{IBIS line sensitivity}.}
\label{sens}
\end{figure}

\begin{figure}[ht!] 
\centerline{\epsfig{file=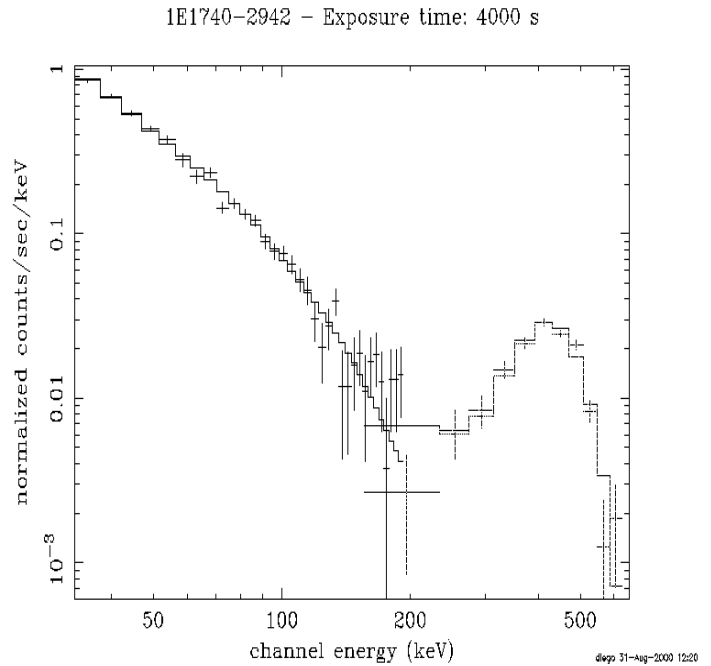,height=3.8in,width=3.8in,angle=-90}}
\vspace{10pt}
\caption{4 ks IBIS simulation of 1E1740-2942.}
\label{1E}
\end{figure}

We have performed simulations with different exposure times and
spectral parameters to determine the sensitivity to
detect an emission line at $\sim$ 500 keV over a
continuum. Figure \ref{sens} shows the minimum exposure time that is
required for a $\sim$3 $\sigma$ detection of such a  feature as a
function of the line flux and width ($\sigma$). The continuum used is that
of Nova Muscae (see Table \ref{table 1}). A feature like that observed in
1E 1740-2942 during October 13-14 1990 with the SIGMA telescope
(Bouchet et al. 1991), for example, can already be detected
with an exposure time of $\sim$250 s. We have also performed some
simulations of this object. The parameters, that we have used,
were a Gaussian line (center = 480 keV, $\sigma$ =
$\sim$100 keV, flux = 1.3$\times$$10^{-2}$ ph cm$^{-2}$ s$^{-1}$) over a
comptonization continuum (kT = 27 keV, $\tau$ = 3.2). In Figure \ref{1E}
we present a 4 ks simulated IBIS spectrum of  1E1740-2942.

\section{Conclusions}

Of course the matrices presented here, being obtained trough
simulations and not by real calibration data, are intrinsically
limited in their accuracy owing to the simplifications inherent
to the simulator programs.
Nevertheless they can be used to analyze
simulated data and to obtain an estimate of the observation
time required to achieve specific scientific objectives.

Future work will include the following effects that are not yet
implemented in the current version:\\

1) Charge loss in ISGRI, caused by the different mobility of the
charge/hole pairs in semiconductors;\\
2) Pixel disuniformity and/or gain variations;\\
3) Multiple interactions in PICsIT;\\
4) Angular dependence of the spectral response;\\
5) Background spatial disuniformity;\\

\end{document}